\documentclass[10pt,conference]{IEEEtran}
\usepackage{amssymb,amsmath,amsfonts,bm}
\usepackage{graphicx}
\begin{document}

\title{Loop Calculus and Belief Propagation for $q$-ary Alphabet: Loop Tower}

\author{
\authorblockN{Vladimir Y. Chernyak}
\authorblockA{Department of Chemistry,
Wayne State University\\ 5101 Cass Ave Detroit, MI 48202\\
{\tt\small chernyak@chem.wayne.edu}} \and
\authorblockN{Michael Chertkov}
\authorblockA{Theoretical Division, T-13 and Center for Nonlinear
Studies,\\ LANL, MS B213, T-13, Los Alamos, NM 87545\\ {\tt\small
chertkov@lanl.gov}} }

\maketitle

\begin{abstract}
Loop Calculus  introduced in \cite{06CCa,06CCb} constitutes a new theoretical tool that explicitly
expresses the symbol Maximum-A-Posteriori (MAP) solution of a general statistical inference problem
via a solution of the Belief Propagation (BP) equations. This finding brought a new significance to
the BP concept, which in the past was thought of as just a loop-free approximation. In this paper
we continue a discussion of the Loop Calculus. We introduce an invariant formulation which allows
to generalize the Loop Calculus approach to a q-are alphabet.
\end{abstract}

The manuscript is organized as follows. In Section \ref{sec:bin} we introduce a new formulation of
the Loop Calculus in terms of a set of gauge transformations that keep the partition function of
the problem invariant. The full expression contains two terms referred to as the ``ground state"
and ``excited states" contributions. The BP equations are interpreted as a special (BP) gauge
fixing condition that emerges as a special orthogonality constraint between the ground and the
excited states. Stated differently, it selects the generalized loop contributions as the only ones
that survive among the excited states. In Section \ref{sec:q} we demonstrate how the invariant
interpretation of the Loop Calculus, introduced in the Section I, allows a natural extension to the
case of a general $q$-ary alphabet. This is achieved via a loop tower sequential construction. The
ground level in the tower is exactly equivalent to assigning one color (out of $q$ available) to
the ``ground state" and considering all ``excited" states to be colored in the remaining $(q-1)$
colors, according to the loop calculus rule. Sequentially, the second level in the tower
corresponds to selecting a loop from the previous step, colored in $(q-1)$ colors, and repeating
the same ground vs excited states partitioning procedure into one and the remaining $(q-2)$ colors,
respectively. The construction proceeds until the complete set of $(q-1)$ levels in the loop tower
(including the corresponding contributions to the partition function) is established. In Section
\ref{sec:free} we discuss an ultimate relation between the loop calculus and the Bethe free energy
variational approach of \cite{05YFW}.

We start with defining a statistical inference problem using the
so-called Forney-style graphical model formulation
\cite{01For,01Loe}. The basic graph, ${C_0}=({\cal V}_0,{\cal
E}_0)$, is described in terms of vertices, ${\cal V}_0=\{a\}$ and
edges, ${\cal E}_0=\{(ab)\}$. Variables, associated with the edges,
assume their values in a $q$-ary alphabet,
$\sigma_{ab}=\sigma_{ba}=0,\cdots,(q-1)$. The probability of a given
configuration of variables ${\bm\sigma}=\{\sigma_{ab}| (ab)\in {\cal
E}_0\}$ on the entire graph is described by
\begin{equation}
p({\bm\sigma})=Z_{C_0}^{-1}\prod\limits_a
f_a\left({\bm\sigma}_a\right),\quad Z_{C_0}=
\sum_{\bm\sigma}\prod\limits_a
f_a\left({\bm\sigma}_a\right),\label{psigma}
\end{equation}
where $Z_{C_0}$ is the normalization coefficient, also known as the
partition function; $f_a({\bm\sigma}_a)$ is an arbitrary positive
function of the variables, ${\bm\sigma}_a\equiv\{\sigma_{ab}|b\in
a,{\cal E}_0\}$, associated with all edges attached to vertex $a$.
$b\in a$ (or conversely $a\in b$) indicates that the vertices $b$
and $a$ share an actual edge of the graph, $(ab)\in {\cal E}_0$. The
marginal probabilities, e.g. associated with edges and vertices,
\begin{eqnarray}
p_a({\bm\sigma}_a)\equiv\sum_{{\bm\sigma}\setminus{\bm\sigma}_{a}}p({\bm\sigma}),
\quad
p_{ab}(\sigma_{ab})\equiv\sum_{{\bm\sigma}\setminus\sigma_{ab}}p({\bm\sigma}),
\label{b-p-ab}
\end{eqnarray}
constitute what one normally needs to evaluate in order to solve a
statistical inference problem. The marginal probabilities can be
also expressed in terms of derivatives of the so-called equilibrium
free energy, ${\cal F}_{C_{0}}=-\ln Z_{C_0}$, with respect to
relevant parameters of the factor functions.

\section{Gauge-Invariant Formulation of Loop Calculus}
\label{sec:bin}

Formally, loop calculus suggests an explicit decomposition of the partition function $Z_{C_0}$ in
terms of a sum over certain loops on the graph ${C_0}$. Below we re-derive the loop calculus in
more general terms compared to \cite{06CCa,06CCb}.

We start with an observation that the partition function, $Z_{C_0}$, is invariant with respect to a
group of linear gauge transformations of the factor functions
\begin{eqnarray}
 f_a({\bm\sigma}_a=(\sigma_{ab},\cdots))\to
 \sum_{\sigma'_{ab}}G_{ab}\left(\sigma_{ab},\sigma'_{ab}\right)
 f_a(\sigma'_{ab},\cdots),\label{fatrans}
\end{eqnarray}
described by
$\hat{G}=\{G_{ab}(\sigma_{ab},\sigma'_{ab});(ab)\in{\cal E}_0\}$
provided the pairs of conjugated matrices $G_{ab}$ and $G_{ba}$ are
related to each other by the special constraint
\begin{eqnarray}
\sum_{\sigma_{ab}}
G_{ab}(\sigma_{ab},\sigma')G_{ba}(\sigma_{ab},\sigma'')=
\delta(\sigma',\sigma''), \label{GG}
\end{eqnarray}
where $\delta(x,y)$ is $1$ if $x=y$ and $0$, otherwise. Except as prescribed by Eq.~(\ref{GG}), the
gauges are chosen independently at different edges of the graphs. This local freedom in selecting
$\hat{G}$ is the key to our further analysis of the partition function, $Z_{C_0}$ now expressed as
\begin{eqnarray}
 && Z_{C_0}=\sum\limits_{\bm\sigma}
 \prod_a\left(\sum\limits_{\bm\sigma'_a}f_a({\bm\sigma}'_a)\prod\limits_{b\in a}
 G_{ab}(\sigma_{ab},\sigma'_{ab})\right)\nonumber\\ &&
 \equiv\sum\limits_{\bm\sigma} \bar{p}\{\hat{G}|{\bm\sigma}\}\equiv
 \mbox{Tr}\left(\bar{p}\{\hat{G}|{\bm\sigma}\}\right),\label{ZG}
\end{eqnarray}
where $\sigma_{ab}=\sigma_{ba}$. We will refer to summation over all
allowed configurations of ${\bm\sigma}$ in Eq.~(\ref{ZG}) as
computing a graphic trace: a conventional trace can be considered as
a special case of the graphic trace for a graph that consists of a
single vertex and a single edge. Our next step in evaluation of
Eq.~(\ref{ZG}) is fixing the gauges, which means imposing
constraints on $\hat{G}$ in addition to Eq.~(\ref{GG}).

It is convenient to distinguish a special term in the sum/trace over ${\bm\sigma}$ in
Eq.~(\ref{ZG}) with all $\sigma_{ab}=0$.  We will refer to this term as the ground or,
alternatively, uncolored state (term), while all the other terms in the sum, which contain at least
one edge with $\sigma_{ab}>0$, are called excited (colored) states. Obviously for a general gauge
choice $\hat{G}$ all kinds of excited states, e.g. with only one edge being excited/colored,
provide nonzero contributions to $Z$. Discussing individual terms in the $\sigma$-sum in
Eq.~(\ref{ZG}) we call a vertex colored if at least one edge attached to it is excited/colored.

A BP-gauge corresponds to such a special choice of $\hat{G}$ that makes vanish any contribution in
the $\sigma$-sum in Eq.~(\ref{ZG}) that has at least one vertex with only one attached colored
edge. Stated differently a BP-gauge prohibits loose excited/colored edges at any vertex. Formally
it is expressed as the following set of conditions
\begin{equation}
\sum\limits_{{\bm\sigma'}_a}f_a({\bm\sigma}')G^{(bp)}_{ab}(\sigma_{ab}\neq
0,\sigma'_{ab}) \prod\limits_{c\in a}^{c\neq
b}G^{(bp)}_{ac}(0,\sigma'_{ac})=0, \label{Gcond}
\end{equation}
enforced independently at any vertex of the graph. Combined with the constraints (\ref{GG}),
Eq.~(\ref{Gcond}) can be re-stated in the vector form depending only on the ground state part of
the gauges:
\begin{eqnarray}
G^{(bp)}_{ba}(0,\sigma'_{ab})=\rho_{a}^{-1}
\sum\limits_{{\bm\sigma'}_a\setminus\sigma'_{ab}}f_a({\bm\sigma}')
\prod\limits_{c\in a}^{c\neq b}G^{(bp)}_{ac}(0,\sigma'_{ac}).
\label{G0}
\end{eqnarray}
with
\begin{eqnarray}
\rho_{a}= \sum\limits_{{\bm\sigma'}_a}f_a({\bm\sigma}')
\prod\limits_{c\in a}G^{(bp)}_{ac}(0,\sigma'_{ac}).
\label{define-lambda}
\end{eqnarray}
We can alternatively derive Eq.~(\ref{G0}) for BP gauges using a variational approach. To that end
we introduce a functional
\begin{equation}
{\cal Z}_{0}(\hat{\epsilon})\equiv\bar{p}\{G|{\bm
0}\}=\prod_{a}\rho_{a}({\bm \epsilon}_{a}), \label{Z0}
\end{equation}
where $\epsilon_{ab}(\sigma_{ab})\equiv G_{ab}(0,\sigma_{ab})$, ${\bm
\epsilon}_a\equiv\{\epsilon_{ab}|b\in a,{\cal V}_0\}$, $\hat{
\epsilon}\equiv\{\epsilon_{ab}|(ab)\in {\cal E}_0\}$, and $\rho_{a}({\bm \epsilon}_{a})$ is given
by Eq.~(\ref{define-lambda}) with $G^{(bp)}_{ac}(0)$ replaced by $\epsilon_{ac}$. The conditions
for the stationary points of ${\cal F}_{0}\equiv -\ln{\cal Z}_0$ with respect to $\hat{ \epsilon}$,
under the additional condition, $\sum_\sigma G_{ab}(0,\sigma)G_{ba}(0,\sigma)=1$, 
recovers Eqs.~(\ref{G0}). Note that the functional ${\cal F}_{0}(\hat{\epsilon})$ as well as the BP
equations (\ref{G0},\ref{define-lambda}) possess some remaining irrelevant gauge freedom with
respect to a set of transformations $\epsilon_{ab}\to\kappa_{ab}\epsilon_{ab}$ with
$\kappa_{ab}\kappa_{ba}=1$. Stated differently, the BP equations fix only the relevant part of the
gauge freedom. A connection between the functional ${\cal F}_0(\hat{\epsilon})$ and the variational
Bethe free energy will be established in Section \ref{sec:free}.

\begin{figure} [t]
\includegraphics[width=0.4\textwidth]{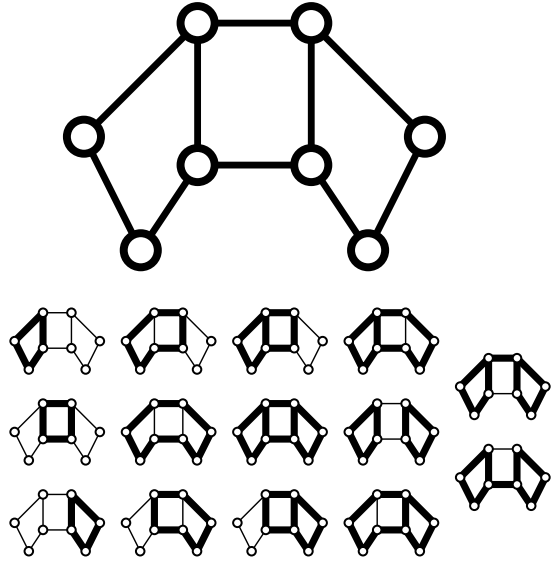}
\caption{Example of a factor graph, ${\it C}_0$  with fourteen
possible generalized loops, $\Omega({\it C}_0)=\{{\it C}_1\}$, shown
in bold on the bottom.} \label{fig:14}
\end{figure}

The conventional form of the BP equations,  in terms of the
``messages" $\eta_{ab}(\sigma_{ab})$,
\begin{eqnarray}
 &&\frac{\exp\left(\eta_{ab}^{(bp)}(\sigma_{ab})+\eta_{ba}^{(bp)}(\sigma_{ab})\right)}{
 \sum_{\sigma_{ab}}\exp\left(\eta_{ab}^{(bp)}(\sigma_{ab})+
 \eta_{ba}^{(bp)}(\sigma_{ab})\right)}\nonumber\\ &&
 =\frac{\sum_{{\bm\sigma}_a\setminus\sigma_{ab}}f_a({\bm\sigma}_a)\exp\left(
 \sum_{b\in a}\eta_{ab}^{(bp)}(\sigma_{ab})\right)}
 {\sum_{{\bm\sigma}_a}f_a({\bm\sigma}_a)\exp\left(
 \sum_{b\in a}\eta_{ab}^{(bp)}(\sigma_{ab})\right)},
 \label{BP-conv}
\end{eqnarray}
is recovered using the following parametrization
\begin{eqnarray} \epsilon_{ab}=G_{ab}(0,\sigma)=\frac{\exp\left(\eta_{ab}(\sigma)\right)}
{\sqrt{\sum_\sigma\exp\left(\eta_{ab}(\sigma)+\eta_{ba}(\sigma)\right)}}.
\label{Geta0}
\end{eqnarray}

Our discussion has been applied so far to the case of a general $q$-ary alphabet. We now turn to
the simplest binary case  $q=2$, where the ground state parametrization (\ref{Geta0}) unambiguously
fixes the excited states: $G_{ab}(1,\sigma)=(1-2\sigma)G_{ba}(0,(\sigma-1)^{2})$. Substituting the
latter expression and Eqs.~(\ref{Gcond},\ref{G0}) into Eq.~(\ref{ZG}) we arrive at the main formula
of the loop calculus for the binary alphabet
\begin{eqnarray}
  Z_{C_0}\!&=&\!Z_{0;C_0}(1\!+\!\sum\limits_{C_1}\!\! r({C_1})),\ \
 r(C_1)\!\equiv\! Z_{0;C_0}^{-1}\bar{p}(G|{\bm\sigma}_{C_1}),
 \nonumber\\
 Z_{0;C_0}&\equiv& \bar{p}(G|{\bm\sigma}_0),\quad
 {\bm\sigma}_0\equiv\{\sigma_{ab}=0|\ \ (ab)\in{C_0}\},
 \nonumber\\ 
 {\bm\sigma}_{C_1}&\equiv&\left\{\begin{array}{cc}
 \sigma_{ab}=1| & \quad (ab)\in {C_1}\\
 \sigma_{ab}=0| & \quad (ab)\in{C_0}\setminus {C_1}.\end{array}\right\}.
 \label{binary}
\end{eqnarray}
where $\{{C_1}\}={\it \Omega}(C_0)$ is the set of generalized loops on the graph, defined as
subgraphs of ${C_0}$ without loose ends,  i.e. with degree of connectivity at any vertex (within
the subgraph) being two or larger.

Beliefs are defined here as substitutes for the exact marginal probabilities (\ref{b-p-ab})
truncated at the first,  ground state, term
\begin{eqnarray}
 b^{(bp)}_{ab}(\sigma_{ab})=G_{ab}^{(bp)}(0,\sigma_{ab})G_{ba}^{(bp)}(0,\sigma_{ab}),
\label{b_ab}\\
b^{(bp)}_a({\bm\sigma}_a)=
 \frac{f_a({\bm\sigma}_a)\prod_{b\in a}G_{ab}^{(bp)}(0,\sigma_{ab})}{
 \sum_{{\bm\sigma}_a}f_a({\bm\sigma}_a)
 \prod_{b\in a}G_{ab}^{(bp)}(0,\sigma_{ab})}.
  \label{b_ab}
\end{eqnarray}
Then a single generalized loop contribution, $r_{C_1}$, is expressed
in terms of the ground state beliefs in the following simple way
\begin{eqnarray}
  r(C_1)\!\!&=&\!\!\!\frac{\prod\limits_{a\in {C_1}}\mu_a}{\prod\limits_{(ab)\in {C_1}}(1-m_{ab}^2)},\
 m_{ab}\!\equiv\!\sum\limits_{\sigma_{ab}}(1\!-\!2\sigma_{ab})b_{ab}^{(bp)}(\sigma_{ab}),
 \nonumber\\ 
 \mu_a & \equiv & \sum_{{\bm\sigma}_a}\left(\prod_{b\in
 a,{C_1}}(1\!-\!2\sigma_{ab}\!-\!m_{ab})\right)b_a^{(bp)}({\bm\sigma}_a).
 \nonumber 
\end{eqnarray}
The loop calculus construction for a simple example is illustrated schematically in
Fig.~\ref{fig:14}.

\section{Loop Tower for $q$-ary alphabet}
\label{sec:q}

Turning to the general $q$-ary alphabet case we first notice that
all considerations and formulas of the introduction and the first
part of Section \ref{sec:bin}, all the way up to Eq.~(\ref{G0}),
actually apply to the general $q$-ary case. Partitioning the
sum/trace over ${\bm\sigma}$ in Eq.~(\ref{ZG}) into the ground-state
term, with ${\bm\sigma}_0$, and the remaining excited-state terms
$\{{\bm\sigma}\setminus{\bm\sigma}_0\}$, and emergence of the
self-consistent set of equations for the ground state gauges
(\ref{Geta0}) are important general features of the gauge fixing
construction. Of course, all the preceding formulas should be
understood in terms of the edge variables that assume values from
$\{0,\cdots,q-1\}$.  Generalization of Eq.~(\ref{binary}) to a
general $q$-ary alphabet reads
\begin{equation}
Z_{C_0} \!=\! Z_{0;{C_0}}\!+\!\!\!\!\!\sum\limits_{C_1\in {\it \Omega}({C_0})}\!\!\!\!\!\!\!
Z_{C_1},\ Z_{C_1}\!=\!\sum_{{\bm\sigma}_{C_1}}\! {\bar p}(G^{(bp)}|{\bm \sigma}_{C_1}).\label{LC1}
\end{equation}
The additional summation over the colored/excited ${\bm \sigma}_{C_1}$ in Eq.~(\ref{LC1}) is a
consequence of the fact that for $q>2$, ${\bm \sigma}_{C_1}$, is not fixed unambiguously, but
rather represents summation over the reduced $(q-1)$-colors rich set, $1,\cdots,q-1$. The BP-gauges
for the original graphical model are described by Eqs.~(\ref{GG},\ref{Gcond}). The set of excited
states gets larger in the $q$-ary case and, consequently, there is a big freedom in selecting the
orthogonal basis set of excited gauges. Selecting one such solution of Eqs.~(\ref{GG},\ref{Gcond}),
$\{G_{ab;{C_0}}^{(bp)}(\sigma_{ab},\sigma'_{ab});(ab)\in{C_0}\}$, and substituting it in
Eq.~(\ref{LC1}) we find that $Z_{C_1}$ becomes the partition function of a reduced graphical model,
defined on a subset $C_1\subset{C_0}$ of ${C_0}$,
\begin{eqnarray}
 && Z_{C_1}=\sum_{{\bm\sigma}_{C_1}} \prod\limits_{a\in C_1}
 f_{1;a}({\bm\sigma}_{a;C_1}),\quad f_{1;a}({\bm\sigma}_{a;C_1})=
 \label{Zc1}\\ &&
 =\sum\limits_{{\bm \sigma}'_a} f_a({\bm\sigma}'_a)
 \prod\limits_{b\in a,{C_0}}G_{ab;C_0}^{(bp)}(\sigma_{ab},\sigma'_{ab})
 \prod\limits_{b\in a,{C_0}}^{b\notin C_1}\delta(\sigma_{ab},0).
 \nonumber
\end{eqnarray}
Here ${\bm\sigma}_{a;C_1}$ is the vector constructed of $\sigma_{ab}$ with $b\in C_1$, with the
components labeled by $\{1,\cdots,q-1\}$. $Z_{C_1}$ may be understood as a partition function of a
reduced graphical model, defined on the graph $C_1$ in terms of a reduced (one element shorter)
alphabet and with the factor functions $f_{1;a}$.

\begin{figure} [t]
\includegraphics[width=0.5\textwidth]{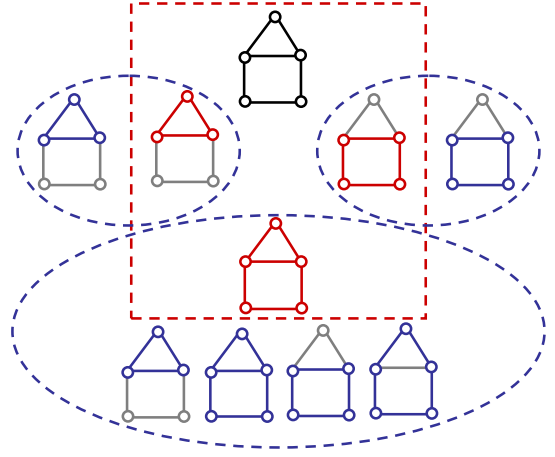}
\caption{Example of a loop tower construction for three colors,
$q=0,1,2$, shown in the Figure in black, red and blue respectively.
First layer of the tower is bounded by the red dashed box,  with the
original graph, ${\cal C}_0$, shown in black and three generalized
loops, $\{{\it C}_1\}=\Omega({\it C}_0)$, shown in red. On the
second layer of the tower each graph from $\Omega({\it C}_0)$
generates its own set of generalized loops. The next layer of
generalized loops, shown in blue, are bounded by three dashed blue
boxes with red graph in a box showing respective element of $\{{\it
C}_1\}$. } \label{fig:tower}
\end{figure}

This reformulation of the partition function of the original problem
in terms of a sum of partition functions over reduced graphical
problems can be repeated sequentially: $Z_{C_0}\to Z_{C_1}\to
Z_{C_2}\to\cdots Z_{C_{q-1}}$ where ${C_0}\supset
C_1\supset\cdots\supset C_{q-2}$ is the tower of loops and $Z_{C_j}$
is the partition function of the graphical model defined in terms of
a $(q-j)$-ary variables on the graph $C_j$, which is a generalized
loop of $C_{j-1}$.  All together one arrives at Eq.~(\ref{LC1})
supplemented by the sequence
\begin{equation}
j\!=\!1,\!\cdots\!,q-2:\ \ Z_{C_j}\!=\!Z_{0;C_j}\!+\!\!\!\!\!\!\sum\limits_{C_{j+1}\in {\it
\Omega}(C_j)}\!\!\!\!\!\! Z_{C_{j+1}}. \label{Zseq}
\end{equation}
Generalization of Eq.~(\ref{Zc1}) becomes
\begin{eqnarray}
 && Z_{C_j}=\sum_{{\bm\sigma}_{C_j}} \prod\limits_{a\in C_j}
 f_{j;a}({\bm\sigma}_{a;C_j}),\label{Zcj}\\
 && f_{j;a}({\bm\sigma}_{a;C_j})
 =\sum\limits_{{\bm \sigma}'_{a;C_{j-1}}}
 f_{j-1;a}({\bm\sigma}'_{a;C_{j-1}})\nonumber\\ &&
 \times\prod\limits_{b\in a,{C_{j-1}}}
 G_{ab;C_{j-1}}^{(bp)}(\sigma_{ab},\sigma'_{ab})
 \prod\limits_{b\in a,{C_{j-1}}}^{b\notin
 C_j}\delta(\sigma_{ab},j-1),
 \nonumber
\end{eqnarray}
where ${\bm\sigma}_{C_j}$ is a vector constructed out of the variables defined on all edges of the
graph $C_j$ with the components labeled by $\{j,\cdots,q-1\}$. The BP gauges in Eqs.~(\ref{Zcj}),
$G_{ab;C_j}^{(bp)}$, are solutions of Eqs.~(\ref{GG},\ref{Gcond}) with the original factor
functions, $f=f_0$ replaced by $f_j$.

\section{Relation to the Bethe free energy approach}
\label{sec:free}

It is known that the exact (equilibrium) free energy of any classical statistical model can be
obtained from a variational principle based on an exact non-equilibrium variational functional of
the full belief, $b({\bm\sigma})$,
\begin{eqnarray}
 {\cal F}_{exact}\{b({\bm\sigma})\}=
 \sum_{\bm\sigma}b({\bm\sigma})\ln\frac{b({\bm\sigma})}{\prod_a
 f_a({\bm\sigma}_a)}.
\label{F_ex}
\end{eqnarray}
The only stationary point of the functional under the normalization condition
\begin{equation}
\sum_{\bm\sigma}b({\bm\sigma})=1, \label{norm}
\end{equation}
reproduces the probability distribution
$b({\bm\sigma})=p({\bm\sigma})$, where $p({\bm\sigma})$ is defined
by Eq.~(\ref{psigma}). This stationary point is actually a minimum.
The value of the exact variational functional at its minimum is
equal to the exact free energy
\begin{equation}
{\cal F}_{exact}\{p({\bm\sigma})\}=F=-\ln Z,\label{Fexact}
\end{equation}
where $Z=Z_{C_0}$, the latter defined above by Eq.~(\ref{psigma}). Hereafter we skip the graph,
$C_0$, index to simplify notations.

Introducing an approximate variational ansatz
\begin{equation}
 b({\bm\sigma})\approx
 \frac{\prod_a b_a({\bm\sigma}_a)}{\prod_{(ab)}b_{ab}(\sigma_{ab})},
\label{b-ans}
\end{equation}
where $b_a$ and $b_{ab}$ are approximations for the corresponding (exact) marginal probabilities,
we substitute it (in the spirit of \cite{05YFW}) into Eq.~(\ref{F_ex}). We further invoke another
approximation (both approximations are actually exact in the case of a tree, i.e. a graph with no
loops)
\begin{eqnarray}
 b_{a}({\bm\sigma}_{a})\approx \sum_{{\bm\sigma}\setminus{\bm\sigma}_{a}}b({\bm\sigma}), \;\;\;
 b_{ab}({\bm\sigma}_{ab})\approx \sum_{{\bm\sigma}\setminus{\bm\sigma}_{ab}}b({\bm\sigma}). \label{approx-2}
\end{eqnarray}
This results in the so-called Bethe (approximate) free energy functional of beliefs $b_a({\bm
\sigma}_a), b_{ac}(\sigma_{ac})$:
\begin{eqnarray}
 && \Phi_{Bethe}=\sum_a \sum_{{\bm \sigma}_a}b_a({\bm\sigma}_a)
 \ln\left(\frac{b_a({\bm\sigma}_a)}{f_a({\bm\sigma}_a)}\right)\nonumber\\ &&
 -\sum_{(a b)}
 \sum_{\sigma_{ab}}b_{ab}(\sigma_{ab})\ln b_{ab}(\sigma_{ab}).
 \label{Bethe}
\end{eqnarray}
We require the beliefs to obey the positivity, normalizability and compatibility  constraints, the
features borrowed from the corresponding exact probabilities given by Eqs.~(\ref{b-p-ab}). Thus, we
have $\forall\ a,c;\ c\in a$ (and inversely $a\in c$):
\begin{eqnarray}
 &&
 0\leq b_a({\bm \sigma}_a),b_{ac}(\sigma_{ac})\leq 1,\label{ineq}\\
 && \sum_{{\bm \sigma}_a} b_a({\bm \sigma}_a)=
 1,\quad \sum_{ \sigma_{ab}} b_{ab}(\sigma_{ab})=
 1,\label{norm}\\
 &&  b_{ac}(\sigma_{ac})\!=\!\!\!\!\!
 \sum_{{\bm \sigma}_a\setminus\sigma_{ac}}\!\!\! b_a({\bm
 \sigma}_a),\ \ b_{ac}(\sigma_{ca})\!=\!\!\!\!\!
 \sum_{\sigma_c\setminus\sigma_{ca}}\!\!\! b_c({\bm \sigma}_c).\label{cons}
\end{eqnarray}

To establish a connection between the Bethe free energy and the
functional ${\cal F}_{0}$ we introduce the effective Lagrangian
\begin{eqnarray}
&& {\cal L}_{Bethe}=\sum_a \sum_{{\bm \sigma}_a}b_a({\bm\sigma}_a)
 \ln\left(\frac{b_a({\bm\sigma}_a)}{f_a({\bm\sigma}_a)}\right)\nonumber\\ &&
 -\sum_{(a b)}
 \sum_{\sigma_{ab}}b_{ab}(\sigma_{ab})\ln b_{ab}(\sigma_{ab})\nonumber\\ &&
 +\sum\limits_{(ab)}\Biggl(\sum\limits_{\sigma_{ab}}
 \ln(\varepsilon_{ab}(\sigma_{ab}))\biggl(b_{ab}(\sigma_{ab})-
 \sum\limits_{{\bm\sigma}_a\setminus\sigma_{ab}}b_a({\bm\sigma}_a)\biggr)
 \nonumber\\ &&
 +\sum\limits_{\sigma_{ba}}
 \ln(\varepsilon_{ba}(\sigma_{ba}))\biggl(b_{ab}(\sigma_{ba})-
 \sum\limits_{{\bm\sigma}_b\setminus\sigma_{ba}}b_b({\bm\sigma}_b)\biggr)\Biggr),
 \label{Lagr}
\end{eqnarray}
that depends on all beliefs that satisfy the normalization
constrains (\ref{norm}) with no constraints on
$\varepsilon_{ab}(\sigma_{ab})$. Requiring vanishing of the
variation with respect to $\varepsilon_{ab}$ obviously leads to the
constraints given by Eq.~(\ref{cons}), and once all constraints are
fulfilled the functional does not depend on $\varepsilon_{ab}$
(which should be considered as gauge symmetry) and coincides with
$\Phi_{Bethe}$ as a function of the beliefs. This implies a
one-to-one correspondence between the extrema of ${\cal L}_{Bethe}$
and Bethe free energy $\Phi_{Bethe}$.

Finding extrema of ${\cal L}_{Bethe}$  with respect to the beliefs
(this can be technically achieved by introducing Lagrange
multipliers for the set of constrains (\ref{norm})) leads to beliefs
that depend explicitly on
$\hat{\varepsilon}\equiv\{\varepsilon_{ab}|(ab)\in {\cal E}_0\}$:
\begin{eqnarray}
 && b_a^{(*)}({\bm\sigma}_a)= (\varrho_{a}({\bm \varepsilon}_{a}))^{-1}f_a({\bm \sigma}_a)\prod\limits_{b\in
 a}\varepsilon_{ab}(\sigma_{ab}) \label{ba*}\\
 && b_{ab}^{(*)}(\sigma_{ab})=\varrho_{ab}^{-1}(\varepsilon_{ab},\varepsilon_{ba})\varepsilon_{ab}(\sigma_{ab})
 \varepsilon_{ba}(\sigma_{ab}),
 \label{bab*}\\
 && {\it\varrho}_{a}({\bm \varepsilon}_{a})\equiv \sum_{{\bm\sigma}_{a}}f_{a}({\bm
 \varepsilon}_{a})\prod_{c\in a}
 \varepsilon_{ac}(\sigma_{ac}), \label{define-e-a} \\
 && \varrho_{ab}(\varepsilon_{ab},\varepsilon_{ba})\equiv
 \sum_{\sigma_{ab}}\varepsilon_{ab}(\sigma_{ab})\varepsilon_{ba}(\sigma_{ab}), \label{define-e-ab}
\end{eqnarray}
where ${\bm \varepsilon}_a=\{\varepsilon_{ab};a\in b,{\cal V}_0\}$.
Substituting the values of beliefs given by
Eqs.~(\ref{ba*},\ref{bab*}) into Eq.~(\ref{Lagr}) results in a
functional that depends on the $\hat{\varepsilon}$ variables only
\begin{eqnarray}
 {\cal F}_{B}(\hat{\varepsilon})=-\sum_{a}\ln\varrho_{a}({\bm \varepsilon}_{a})+\sum_{(ab)}
 \ln\left(\varrho_{ab}(\varepsilon_{ab},\varepsilon_{ba}))\right).
 \label{F-B}
\end{eqnarray}
The functional ${\cal F}_{B}$ possesses strong gauge symmetry: it is
invariant under a set of transformations
$\varepsilon_{ab}\to\kappa_{ab}\varepsilon_{ab}$. The gauge can be
partially fixed by implementing a gauge (normalization) condition
\begin{eqnarray}
 \sum_{\sigma_{ab}}\varepsilon_{ab}(\sigma_{ab})\varepsilon_{ba}(\sigma_{ab})=1.
 \label{gauge-F-B}
\end{eqnarray}
Implementing this constraint, the second term in Eq.~(\ref{F-B}) vanishes. This means that
switching from the notations of Section \ref{sec:bin} to our current notations,
$\epsilon\to\varepsilon$ and $\rho\to\varrho$, we arrive at ${\cal F}_{0}=\cal{F}_{B}$. Stated more
formally, ${\cal F}_{0}$ introduced earlier represents the gauge-invariant functional ${\cal
F}_{B}$ in a particular gauge determined by Eq.~(\ref{gauge-F-B}). This implies a one-to-one
correspondence of the extrema of ${\cal F}_{0}$ to the extrema of ${\cal L}_{Bethe}$, and therefore
to the extrema of the Bethe free energy $\Phi_{Bethe}$.

\section{Discussions and Conclusions}
\label{sec:con}

We first summarize the results presented in the manuscript. We have introduced a group of gauge
transformations that keep the partition function of the graphical model invariant, and naturally
split the gauges into the ``ground" and ``excited" parts.  The partition function is decomposed
into the principal ground and many excited terms. Each excited contribution is interpreted in terms
of an excited subgraph constructed from excited edges. Requiring that only excited subgraphs with
no loose ends contribute to the partition function sets the BP equations for the ground gauges. We
show that the BP equations can be derived using a variational principle for the partition function
as a function of the ground gauges. Further consideration differs for the binary and $q$-ary
alphabets. In the binary case the excited gauges are fixed unambigiously, generating the binary
loop series over generalized loops for the partition function \cite{06CCa,06CCb,06CCc}. In the
$q$-ary case we pick one (of many possible) excited gauges and presenting the full partition
function as a sum over generalized loops. Each contribution labeled by a generalized loop can be
viewed as a new graphical model defined on this loop with a new set of factor functions. The loop
decomposition procedure  is applied again, introducing new ground and excited gauges, fixing the
gauges, etc. The procedure repeated for $(q-1)$ layers builds a $q$-store loop tower. Finally, we
showed that the BP-gauges can be determined using a variational principle and related the
corresponding functional ${\cal F}_{0}$ to the Bethe free energy functional constructed in the
spirit of \cite{05YFW}.

These results open new venues for further development, and also
raise a set of important and challenging questions listed below. (1)
Already the lowest level BP equations in the loop tower, the ground
BP-gauge may have multiple solutions. Our construction applied to
different solutions will generate different loop decompositions for
the partition function. The question is, whether a preferred
solution is in a way better then the others? Naive intuition
suggests that BP gauge with the highest value of $Z_{C_0}$ would
serve better. (2) Furthermore, in the case of a $q$-ary alphabet
with $q>2$ positivity of the factor functions at higher tower levels
is not guaranteed. The positivity would be desirable for
interpreting the auxiliary graphical problems as some actual
statistical inference problems, with the factor functions related to
probabilities. On the other hand, there is a big freedom in
selecting the excited gauges, and a question surfaces: could one
select the excited gauges in a way to guarantee positivity of the
higher-level factor functions? (3) The BP ground state contribution
to the partition function, $Z_{{\it C}_0}$ is positive by
construction, however the signs of the excited terms can alternate.
This raises a couple of important questions. How do the signs of the
loop terms depend on the factor functions and the graphical model
itself? Based on our previous experiments \cite{06CCc}, we know that
emergence of an excited loop contribution comparable to the ground
state alerts for a possible failure of BP as an approximation to
exact inference. How exactly does the sign alternation and relative
value of the tower loop contributions affect success or failure of
BP as an approximation? (4) The equilibrium Bethe free energy
estimates the value of the partition function, however the
variational derivation sketched in Section \ref{sec:free} does not
guarantee that the resulting ${\cal F}_0$ is actually larger then
the exact ${\cal F}$. Indeed in the transition from
Eqs.~(\ref{F_ex},\ref{b-ans},\ref{approx-2}) to Eqs.~(\ref{Bethe})
we further discuss the latter formulation completely ignoring the
fact that the conditions (\ref{approx-2}) can be violated for the
resulting BP solutions. How does this violation affect the relation,
${\cal F}\gtrless{\cal F}_0$, and what are the consequences of this
inequality for the loop series?

We conclude with mentioning some future research directions. As
demonstrated in \cite{06CCc}, the loop calculus is suggestive of an
efficient truncation of the full series that can potentially improve
the BP approximation. This idea can also be extended to the $q$-ary
alphabet case,  with the tower truncated at some relatively low
level. This approach can obviously find interesting application in
decoding of non-binary codes and also in problems, such as computer
vision, that require a multi-valued data reconstruction. The loop
tower approach can also be extended to the analogous case of
continuous alphabet. In this case the ground state gauges satisfy a
set of integral equations, while the ground and excited states that
define the gauges become elements of functional infinite-dimensional
Hilbert spaces, which makes the tower heights unlimited and the
tower loop decomposition turns into an infinite series. Finally, we
note that the gauge conditions may be chosen in some other non-BP
way. BP-gauge is of a special importance for dilute locally
tree-like graphs simply because in the loop-free case the whole loop
hierarchy (the entire loop tower) disappears. One could conjecture
that for some other classes of graphical models, e.g. those
naturally defined on regular lattices, similar cancelations can take
place for some alternative specially selected gauges.

The work at Los Alamos was carried out under the auspices of the National Nuclear Security
Administration of the U.S. Department of Energy at Los Alamos National Laboratory under Contract
No. DE-AC52-06NA25396. VYC also acknowledges the support through the start-up grant from Wayne
State University.

\end{document}